\documentclass[runningheads,a4paper]{llncs}

\usepackage{graphicx}
\usepackage{amsmath,amsfonts,amscd,amssymb}
\usepackage{dsfont}
\renewcommand{\vec}[1]{\mathbf{#1}}
\usepackage{hyperref}
\hypersetup{pdfauthor={Felix Biessmann},pdftitle={Political Bias Prediction}}

\newcommand{\R}{\mathds{R}}
\usepackage{multicol}
\usepackage{multirow}
\usepackage{pbox}
\usepackage{url}

\begin{document}

\mainmatter  

\title{Automating Political Bias Prediction}

\titlerunning{Political Bias Prediction}

%
%
\author{
Felix Biessmann
\thanks{felix.biessmann@gmail.com}
\authorrunning{Political Bias Prediction}}


%
%

\toctitle{Political Bias Prediction}
\tocauthor{}
\maketitle

\begin{abstract} 
Every day media generate large amounts of text. An unbiased view on media reports requires an understanding of the political bias of media content. Assistive technology for estimating the political bias of texts can be helpful in this context. This study proposes a simple statistical learning approach to predict political bias from text. Standard text features extracted from speeches and manifestos of political parties are used to predict political bias in terms of political party affiliation and in terms of political views. Results indicate that political bias can be predicted with above chance accuracy. Mistakes of the model can be interpreted with respect to changes of policies of political actors. Two approaches are presented to make the results more interpretable: a) discriminative text features are related to the political orientation of a party and b) sentiment features of texts are correlated with a measure of political power. Political power appears to be strongly correlated with positive sentiment of a text. To highlight some potential use cases a web application shows how the model can be used for texts for which the political bias is not clear such as news articles.
\end{abstract} 

\section{Introduction}
\label{sec:intro}
Modern media generate a large amount of content at an ever increasing rate. Keeping an unbiased view on what media report on requires to understand the political bias of texts. In many cases it is obvious which political bias an author has. In other cases some expertise is required to judge the political bias of a text. 
When dealing with large amounts of text however there are simply not enough experts to examine all possible sources and publications. Assistive technology can help in this context to try and obtain a more unbiased sample of information. 

Ideally one would choose for each topic a sample of reports from the entire political spectrum in order to form an unbiased opinion. But ordering media content with respect to the political spectrum at scale requires automated prediction of political bias. The aim of this study is to provide empirical evidence  indicating that leveraging open data sources of german texts, automated political bias prediction is possible with above chance accuracy. These experimental results  confirm and extend previous findings \cite{Yu2008,Hirst2014}; a novel contribution of this work is a proof of concept which applies this technology to sort news article recommendations according to their political bias. 

When human experts determine political bias of texts they will take responsibility for what they say about a text, and they can explain their decisions. This is a key difference to many statistical learning approaches. Not only is the responsibility question problematic, it can also be difficult to interpret some of the decisions. In order to validate and explain the predictions of the models three strategies that allow for better interpretations of the models are proposed. First the model misclassifications are related to changes in party policies. Second univariate measures of correlation between text features and party affiliation allow to relate the predictions to the kind of information that political experts use for interpreting texts. Third sentiment analysis is used to investigate whether this aspect of language has discriminatory power. 

In the following \autoref{sec:related} briefly surveys some related work, thereafter \autoref{sec:data} gives an overview of the data acquisition and preprocessing methods, \autoref{sec:model} presents the model, training and evaluation procedures; in \autoref{sec:results} the results are discussed and \autoref{sec:conclusion} concludes with some interpretations of the results and future research directions. 

\section{Related Work}\label{sec:related}
Throughout the last years automated content analyses for political texts have been conducted on a variety of text data sources (parliament data \, blogs, tweets, news articles, party manifestos) with a variety of methods, including sentiment analysis, stylistic analyses, standard bag-of-word (BOW) text feature classifiers and more advanced natural language processing tools. 
While a complete overview is beyond the scope of this work, the following paragraphs list similarities and differences between this study and previous work. For a more complete overview we refer the reader to \cite{Grimmer2013,Kaal2014}. 

A similar approach to the one presented here was taken in \cite{Yu2008}. The authors extracted BOW feature vectors and applied linear classifiers to predict political party affiliation of US congress speeches. They used data from the two chambers of the US congress, House and Senat, in order to assess generalization performance of a classifier trained on data from one chamber and tested on data from another. They found that accuracies of the model when trained on one domain and tested on another were significantly decreased. Generalization was also affected by the time difference between the political speeches used for training and those used for testing. 

Other work has focused on developing dedicated methods for predicting political bias. Two popular methods are WordFish \cite{Slapin08ascaling} and WordScores \cite{Laver2003}, or improved versions thereof, see e.g. \cite{Lowe09scalingpolicy}. These approaches have been very valuable for {\em a posteriori} analysis of historical data but they do not seem to be used as much for analyses of new data in a predictive analytics setting. Moreover direct comparisons of the results obtained with these so called {\em scaling} methods with the results of the present study or those of studies as \cite{Yu2008} are difficult, due to the different modeling and evaluation approaches: Validations of WordFish/WordScore based analyses often compare parameter estimates of the different models rather than predictions of these models on held-out data with respect to the same type of labels used to train the models. 

Finally Hirst et al conducted a large number of experiments on data from the Canadian parliament and the European parliament; these experiments can be directly compared to the present study both in terms of methodology but also with respect to their results \cite{Hirst2014}. The authors show that a linear classifier trained on parliament speeches uses language elements of defense and attack to classify speeches, rather than ideological vocabulary. The authors also argue that emotional content plays an important role in automatic analysis of political texts. Furthermore their results show a clear dependency between length of a political text and the accuracy with which it can be classified correctly. 

Taken together, there is a large body of literature in this expanding field in which scientists from quantitative empirical disciplines as well as political science experts collaborate on the challenging topic of automated analysis of political texts. Except for few exceptions most previous work has focused on binary classification\footnote{Many parliaments only have two parties and many studies chose binary classification schemes within manually defined topics and generic schemes such as conservative vs liberal or left vs right.} or on assignment of a one dimensional policy position (mostly left vs right). Yet many applications require to take into account more subtle differences in political policies. This work focuses on more fine grained political view prediction: for one, the case of the german parliament is more diverse than two parliament systems, allowing for a distinction between more policies; second the political view labels considered are more fine grained than in previous studies. While previous studies used such labels only for partitioning training data \cite{Slapin08ascaling} (which is not possible at test time in real-world applications where these labels are not known) the experiments presented in this study directly predict these labels. Another important contribution of this work is that many existing studies are primarily concerned with {\em a posteriori} analysis of historical data. This work aims at prediction of political bias on out-of-domain data with a focus on the practical application of the model on new data, for which a prototypical web application is provided. The experiments on out-of-domain generalization complement the work of \cite{Yu2008,Hirst2014} with results from data of the german parliament and novel sentiment analyses. 

\section{Data Sets and Feature Extraction}\label{sec:data}
All experiments were run on publicly available data sets of german political texts and standard libraries for processing the text. The following sections describe the details of data acquisition and feature extraction. 

\subsection{Data}
Annotated political text data was obtained from two sources: a) the discussions and speeches held in the german parliament ({\em Bundestag}) and b) all manifesto texts of parties running for election in the german parliament in the current 18th and the last, 17th, legislation period.

\paragraph{Parliament discussion data} Parliament texts are annotated with the respective party label, which we take here as a proxy for political bias. The texts of parliament protocols are available through the website of the german bundestag\footnote{\url{https://www.bundestag.de/protokolle}}; an open source API was used to query the data in a cleaned and structured format\footnote{\url{https://github.com/bundestag}}. In total 22784 speeches were extracted for the 17th legislative period and 11317 speeches for the 18th period, queried until March 2016. 

\paragraph{Party manifesto data}
For party manifestos another openly accessible API was used, provided by the Wissenschaftszentrum Berlin (WZB). The API is released as part of the {\em Manifestoproject} \cite{manifesto}. The data released in this project comprises the complete manifestos for each party that ran for election enriched with annotations by political experts. Each sentence (in some cases also parts of sentences) is annotated with one of 56 political labels. Examples of these labels are {\em pro/contra protectionism, decentralism, centralism, pro/contra welfare}; for a complete list and detailed explanations on how the annotators were instructed see \cite{leftright}. The set of labels was developed by political scientists at the WZB and released for public use. All manifestos of parties that were running for election in this and the last legislative period were obtained. In total this resulted in 29451 political statements that had two types of labels: First the party affiliation of each political statement; this label was used to evaluate the party evaluation classifiers trained on the parliament speeches. For this purpose the data acquisition was constrained to only those parties that were elected into the parliament. Next to the party affiliation the political view labels were extracted. For the analyses based on political view labels all parties were considered, also those that did not make it into the parliament. 

The length of each annotated statement in the party manifestos was rather short. The longest statement was 522 characters long, the 25\%/50\%/75\% percentiles were 63/95/135 characters. Measured in words the longest data point was 65 words and the 25\%/50\%/75\% percentiles were 8/12/17 words, respectively. This can be considered as a very valuable property of the data set, because it allows a fine grained resolution of party manifestos. However for a classifier (as well as for humans) such short sentences can be rather difficult to classify. In order to obtain less 'noisy' data points from each party -- for the party affiliation task only -- all statements were aggregated into political topics using the manifesto code labels. Each political view label is a three digit code, the first digit represents the political domain. In total there were eight political domains (topics): {\em External Relations, Freedom and Democracy, Political System, Economy, Welfare and Quality of Life, Fabric of Society, Social Groups} and a topic {\em undefined}, for a complete list see also \cite{leftright}. These 8 topics were used to aggregate all statements in each manifesto into topics. Most party manifestos covered all eight of them, some party manifestos in the 17th Bundestag only covered seven. 

\subsection{Bag-of-Words Vectorization}\label{sec:bow-vectorization}
First each data set was segmented into semantic units; in the case of parliament discussions this were the speeches, in the case of the party manifesto data semantic units were the sentences or sentence parts associated with one of the 56 political view labels.  Parliament speeches were often interrupted; in this case each uninterrupted part of a speech was considered a semantic unit. Strings of each semantic unit were tokenised and transformed into bag-of-word vectors as implemented in scikit-learn \cite{scikit-learn}. The general idea of bag-of-words vectors is to simply count occurrences of words (or word sequences, also called {\em n-grams}) for each data point. A data point is usually a document, here it is the semantic units of parliament speeches and manifesto sentences, respectively. The text of each semantic unit is transformed into a vector $\vec{x}\in\mathds{R}^d$ where $d$ is the size of the dictionary; the $w$th entry of $\vec{x}$ contains the (normalized) count of the $w$th word (or sequence of words) in our dictionary. Several options for vectorizing the speeches were tried, including term-frequency-inverse-document-frequency normalisation, n-gram patterns up to size $n=3$ and several cutoffs for discarding too frequent and too infrequent words. All of these hyperparameters were subjected to hyperparameter optimization as explained in \autoref{sec:crossvalidation}.

\section{Classification Model and Training Procedure}\label{sec:model}
Bag-of-words feature vectors were used to train a multinomial logistic regression model. Let $y\in\{1,2,\dots,K\}$ be the true  label, where $K$ is the total number of labels and $\vec{W}=[\vec{w}_1,\dots,\vec{w}_K]\in\R^{d\times K}$ is the concatenation of the weight vectors $\vec{w}_k$ associated with the $k$th party then 
\begin{eqnarray}\label{eq:logreg_multiclass}
p(y=k|\vec{x},\vec{W}) = &\frac{e^{z_k}}{\sum_{j=1}^K e^{z_j}} \qquad \textrm{with }  z_k=&\vec{w}_k^{\top}\vec{x} \\\nonumber
\end{eqnarray}
We estimated $\vec{W}$ using quasi-newton gradient descent. The optimization function was obtained by adding a penalization term to the negative log-likelihood of the multinomial logistic regression objective and the optimization hence found the $\vec{W}$ that minimized
\begin{equation}\label{eq:objective}
L(\vec{W}, \vec{x}, \gamma) = - \log{\frac{e^{z_k}}{\sum_{j=1}^K e^{z_j}}}+ \gamma \| \vec{W} \|_{F}
\end{equation}
Where $\|~\|_F$ denotes the Frobenius Norm and $\gamma$ is a regularization parameter controlling the complexity of the model. 
The regularization parameter was optimized on a log-scaled grid from $10^{-4,\dots,4}$. The performance of the model was optimized using the classification accuracy, but we also report all other standard measures, precision ($TP / (FP + TP$), recall ($TP / (TP + FN)$) and f1-score ($2\times (Prec. \times Rec) / (Prec + Rec.)$). \\

Three different classification problems were considered: 
\begin{enumerate}
\item {\bf Classification of party affiliation} (five class / four class problem)
\item {\bf Classification of government membership} (binary problem)
\item {\bf Classification of political views} (56 class problem)
\end{enumerate}

Party affiliation is a five class problem for the 17th legislation period, and a four class problem for the 18th legislation period. Political view classification is based on the labels of the manifesto project, see \autoref{sec:data} and \cite{leftright}. 
For each of first two problems, party affiliation and government membership prediction, classifiers were trained on the parliament speeches. For the third problem classifiers were trained only on the manifesto data for which political view labels were available. 

\subsection{Optimisation of Model Parameters}\label{sec:crossvalidation}
The model pipeline contained a number of  hyperparameters that were optimised using cross-validation.  
We first split the training data into a training data set that was used for optimisation of hyperparameters and an held-out test data set for evaluating how well the model performs on in-domain data; wherever possible the generalisation performance of the models was also evaluated on out-of domain data. Hyperparameters were optimised using grid search and 3-fold cross-validation within the training set only: A cross-validation split was made to obtain train/test data for the grid search and for each setting of hyperparameters the entire pipeline was trained and evaluated -- no data from the in-domain evaluation data or the out-of-domain evaluation data were used for hyperparameter optimisation. For the best setting of all hyperparameters the pipeline was trained again on all training data and evaluated on the evaluation data sets. For party affiliation prediction and government membership prediction the training and test set were 90\% and 10\%, respectively, of all data in a given legislative period. Out-of-domain evaluation data were the texts from party manifestos. For the political view prediction setting there was no out-of-domain evaluation data, so all labeled manifesto sentences in both legislative periods were split into a training and evaluation set of 90\% (train) and 10\% (evaluation). 

\subsection{Sentiment analysis}\label{sec:sentiment_analysis_methods}
A publicly available key word list was used to extract sentiments \cite{remquahey2010}. A sentiment vector $\vec{s}\in\R^d$ was constructed from the sentiment polarity values in the sentiment dictionary. The sentiment index used for attributing positive or negative sentiment to a text was computed as the cosine similarity between BOW vectors $\vec{x}\in\R^d$ and $\vec{s}$

\begin{align}
\frac{\vec{s}^\top \vec{x}}{\|\vec{s}\|\|\vec{x}\|}
\end{align}

\subsection{Analysis of bag-of-words features}\label{sec:correlations_methods}
While interpretability of linear models is often propagated as one of their main advantages, doing so naively without modelling the noise covariances can lead to wrong conclusions, see e.g. \cite{Zien2009,Haufe2013}; interpreting coefficients of linear models (independent of the regularizer used) implicitly assumes uncorrelated features; this assumption is violated by the text data used in this study. Thus direct interpretation of the model coefficients $\vec{W}$ is problematic. In order to allow for better interpretation of the predictions and to assess which features are discriminative correlation coefficients between each word and the party affiliation label were computed. The words corresponding to the top positive and negative correlations are shown in \autoref{sec:word_party_correlations}.

\section{Results}\label{sec:results}

The following sections give an overview of the results for all political bias prediction tasks. Some interpretations of the results are highlighted and a web application of the models is presented at the end of the section.

\subsection{Predicting political party affiliation}
The results for the political party affiliation prediction on held-out parliament data and on evaluation data are listed in \autoref{tab:results_17} for the 17th Bundestag and in \autoref{tab:results_18} for the 18th Bundestag, respectively. 
Shown are the evaluation results for in-domain data (held-out parliament speech texts) as well as the out-of-domain data; the party manifesto out-of-domain predictions were made on the sentence level. 

When predicting party affiliation on text data from the same domain that was used for training the model, average precision and recall values of above 0.6 are obtained. These results are comparable to those of \cite{Hirst2014} who report a classification accuracy of 0.61 on a five class problem of prediction party affiliation in the European parliament; the accuracy for the 17th Bundestag is 0.63, results of the 18th Bundestag are difficult to compare as the number of parties is four and the legislation period is not finished yet.
For out-of domain data the models yield significantly lower precision and recall values between 0.3 and 0.4.  This drop in out of domain prediction accuracy is in line with previous findings \cite{Yu2008}.
A main factor that made the prediction on the out-of-domain prediction task particularly difficult is the short length of the strings to be classified, see also \autoref{sec:data}. In order to investigate whether this low out-of-domain prediction performance was due the domain difference (parliament speech vs manifesto data) or due to the short length of the data points, the manifesto data was aggregated based on the topic. The manifesto code political topics labels were used to concatenate texts of each party to one of eight topics, see \autoref{sec:data}. The topic level results are shown in \autoref{tab:results_topic} and \autoref{tab:confusion_topic} and demonstrate that when the texts to be classified are sufficiently long and the word count statistics are sufficiently dense the classification performance on out of domain data can achieve in the case of some parties reliably precision and recall values close to 1.0. This increase is in line with previous findings on the influence of text length on political bias prediction accuracy \cite{Hirst2014}.

\begin{table}[t]
\caption{
\label{tab:results_17}
Classification performance on the party affiliation prediction problem for data from the 17th legislative period on test set and evaluation set, respectively. Predictions on the manifesto data was done on {\bf sentence level}; $N$ denotes number of data points in the evaluation set.
}
\begin{center}
\begin{tabular}{lcccccccc}
& \multicolumn{4}{c}{\bf Held-out parliament speeches} & \multicolumn{4}{c}{\bf Party Manifestos}\\
    &         precision    &recall &  f1-score  & N    &         precision    &recall &  f1-score  & N\\
\hline \hline
       cducsu   &    0.62  &    0.81  &    0.70  &     706&0.26    &  0.58   &   0.36    &  2030\\
        fdp    &   0.70   &   0.37  &    0.49    &   331&0.38   &   0.28    &  0.33   &   2319\\
     gruene &      0.59  &    0.40   &   0.48   &    298&0.47  &    0.20   &   0.28 &     3747\\
      linke    &   0.71   &   0.61  &    0.65    &   338&0.30   &   0.47   &   0.37    &  1701\\
        spd   &    0.60   &   0.69  &    0.65   &    606&0.26   &   0.16   &   0.20  &    2278\\
\hline
avg / total &      0.64   &   0.63   &   0.62    &  2279 &0.35    &  0.31 &     0.30   &  12075
\end{tabular}
\end{center}
\end{table}

\begin{table}[t]
\caption{
\label{tab:results_18}
Classification performance on the party affiliation prediction problem for data from the 18th legislative period on test set and evaluation set, respectively.  Predictions on the manifesto data was done on {\bf sentence level}.
}
\begin{center}
\begin{tabular}{lcccccccc}
& \multicolumn{4}{c}{\bf Held-out parliament speeches} & \multicolumn{4}{c}{\bf Party Manifestos}\\
    &         precision    &recall &  f1-score  & N    &         precision    &recall &  f1-score  & N\\
\hline \hline
    cducsu    &   0.66   &   0.82   &   0.73    &   456 & 0.32  &    0.64  &    0.43    &  2983\\
     gruene   &    0.68    &  0.54   &   0.60    &   173   &0.59   &   0.15   &   0.23   &   5674\\
      linke     &  0.77  &    0.58    &  0.66    &   173 & 0.36   &   0.48   &   0.41   &   2555\\
        spd     &  0.60  &    0.54   &   0.57    &   330 & 0.26 &     0.31   &   0.28     & 2989\\
\hline
avg / total    &   0.66  &    0.66  &    0.65   &   1132&  0.42 &     0.34  &    0.32&     14201\\
\end{tabular}
\end{center}

\end{table}

\begin{table}[t]
\caption{
\label{tab:results_topic}
{\bf Topic level classification performance} on the party affiliation prediction problem for data from the evaluation set (manifesto texts) of the 17th legislative period. In contrast to single sentence level predictions (see \autoref{tab:results_17}, \autoref{tab:results_18}, \autoref{tab:confusion} for results and \autoref{sec:data} for topic definitions) the predictions made on topic level are reliable in many cases. Note that all manifesto topics of the green party in the 18th Bundestag are predicted to be from the parties of the governing coalition, CDU/CSU or SPD.}
\begin{center}
\begin{tabular}{lcccc}
& \multicolumn{4}{c}{\bf 17th Bundestag} \\
    &         precision    &recall &  f1-score  & N  \\
    \hline
        \hline
cducsu     &  0.64  &    1.00  &    0.78    &     7\\
       fdp    &   1.00    &  1.00    &  1.00    &     7\\
    gruene  &     1.00  &    0.86  &    0.92    &     7\\
     linke    &   1.00   &   1.00     & 1.00    &     7\\
       spd   &    0.80   &   0.50    &  0.62     &    8\\
    \hline
    avg / total  &     0.88   &   0.86   &   0.86  &      36\\
\end{tabular}
\quad
\begin{tabular}{lcccc}
& \multicolumn{4}{c}{\bf 18th Bundestag} \\
    &         precision    &recall &  f1-score  & N  \\
    \hline
    \hline
  cducsu     &  0.50   &   1.00  &    0.67     &    8\\
       gruene  &     0.00   &   0.00   &   0.00  &       8\\
      linke     &  1.00   &   0.88   &   0.93  &       8\\
        spd     &  0.56  &    0.62  &    0.59   &      8\\
\hline
avg / total   &    0.51 &     0.62   &   0.55   &     32\\
\end{tabular}
\end{center}

\end{table}

In order to investigate the errors the models made confusion matrices were extracted for the predictions on the out-of-domain evaluation data for sentence level predictions (see \autoref{tab:confusion}) as well as topic level predictions (see \autoref{tab:confusion_topic}). One example illustrates that the mistakes the model makes can be associated with changes in the party policy. The green party has been promoting policies for renewable energy and against nuclear energy in their manifestos prior to both legislative periods. Yet the statements of the green party are more often predicted to be from the government parties than from the party that originally promoted these green ideas, reflecting the trend that these legislative periods governing parties took over policies from the green party. This effect is even more pronounced in the topic level predictions: a model trained on data from the 18th Bundestag predicts all manifesto topics of the green party to be from one of the parties of the governing coalition, CDU/CSU or SPD. \\

\paragraph{Government membership prediction} Next to the party affiliation labels also government membership labels were used to train models that predict whether or not a text is from a party that belonged to a governing coalition of the Bundestag. In \autoref{tab:results_binary_17} and \autoref{tab:results_binary_18} the results are shown for the 17th and the 18th Bundestag, respectively. While the in-domain evaluation precision and recall values reach values close to 0.9, the out-of-domain evaluation drops again to values between 0.6 and 0.7. This is in line with the results on binary classification of political bias in the Canadian parliament \cite{Yu2008}. The authors report classification accuracies between 0.8 and 0.87, the accuracy in the 17th Bundestag was 0.85. While topic-level predictions were not performed in this binary setting, the party affiliation results in \autoref{tab:results_topic} suggest that a similar increase in out-of-domain prediction accuracy could be achieved when aggregating texts to longer segments. 

\begin{table}[t]\label{tab:conf_mat_four_class}
\caption{\label{tab:confusion} {\bf Confusion matrices (sentence level)} for predictions on evaluation data (party manifestos); classifiers were trained on parliament speeches for the 17th legislative period (left) and 18th legislative period (right); the most prominent effect is the high likelihood for a party to be taken as the strongest, governing party, cdu/csu. This can be interpreted as a change in policies of the conservative party cdu/csu towards the policies of the green party.}
\begin{tabular}{lccccccc}
 \multicolumn{8}{c}{\bf 17th Bundestag}\\
 \\
&&& \multicolumn{5}{c}{Predicted}\\
&&& cducsu & fdp& gruene& linke& spd\\
\hline
\multirow{5}{*}{\rotatebox{90}{\pbox{3cm}{\centering True}}}& &cducsu &1186 &289& 178& 198& 179\\
&&fdp &882& 658& 236& 329& 214\\
&&gruene &1174& 404& 764& 941& 464\\
&&linke &388& 92& 214& 806& 201\\
&&spd &999& 268& 240& 398& 373\\
\end{tabular}
\hfill
\begin{tabular}{lcccccc}
 \multicolumn{7}{c}{\bf 18th Bundestag}\\
 \vspace{1em}\\
&&& \multicolumn{4}{c}{Predicted}\\
&&& cducsu & gruene& linke& spd\\
\hline
\multirow{4}{*}{\rotatebox{90}{\pbox{4.7cm}{\centering True}}}&&cducsu&1912& 156& 331& 584\\
&&gruene&2092& 827& 1311& 1444\\
&&linke&596& 186& 1216& 557\\
&&spd&1284& 226& 563& 916\\
\end{tabular}
\vspace{1em}
\end{table}

\begin{table}[t]\label{tab:conf_mat_four_class}
\caption{\label{tab:confusion_topic} {\bf Confusion matrices (topic level)} for predictions on evaluation data (party manifestos) for classifiers trained on parliament speeches for the 17th legislative period (left) and 18th legislative period (right).}
\begin{tabular}{lccccccc}
 \multicolumn{8}{c}{\bf 17th Bundestag}\\
 \\
&&& \multicolumn{5}{c}{Predicted}\\
&&& cducsu & fdp& gruene& linke& spd\\
\hline
\multirow{5}{*}{\rotatebox{90}{\pbox{3cm}{\centering True}}} &&cducsu &7& 0& 0& 0& 0\\
&&fdp&0& 7& 0& 0& 0\\
&&gruene&0& 0& 6& 0& 1\\
&&linke&0& 0& 0& 7& 0\\
&&spd&4& 0& 0& 0& 4\\
\end{tabular}
\hfill
\begin{tabular}{lcccccc}
 \multicolumn{7}{c}{\bf 18th Bundestag}\\
 \vspace{1em}\\
&&& \multicolumn{4}{c}{Predicted}\\
&&& cducsu & gruene& linke& spd\\
\hline
\multirow{4}{*}{\rotatebox{90}{\pbox{4.7cm}{\centering True}}}&&cducsu&8& 0& 0& 0\\
&&gruene&4& 0& 0& 4\\
&&linke&1& 0& 7& 0\\
&&spd&3& 0& 0& 5\\
\end{tabular}
\vspace{1em}\\
\end{table}

\begin{table}[t]
\caption{
\label{tab:results_binary_17}
Classification performance on the binary prediction problem in the 17th legislative period, categorizing speeches into government (FDP/CDU/CSU) and opposition (Linke, Gr\"une, SPD).
}
\begin{center}
\begin{tabular}{lcccccccc}
& \multicolumn{4}{c}{\bf Held-out parliament speeches} & \multicolumn{4}{c}{\bf Party Manifestos}\\
    &         precision    &recall &  f1-score  & N    &         precision    &recall &  f1-score  & N\\
\hline \hline
government    &   0.83   &   0.84&      0.84     & 1037 & 0.49  &    0.59 &     0.54  &    4349\\
 opposition     &  0.86  &    0.86   &   0.86    &  1242 & 0.74 &     0.66  &    0.70   &   7726\\
\hline
avg / total   &    0.85 &     0.85 &     0.85  &    2279 & 0.65  &     0.63  &     0.64  &    12075\\

\end{tabular}
\end{center}
\end{table}

\begin{table}[t]
\caption{
\label{tab:results_binary_18}
Classification performance on the binary prediction problem in the 18th legislative period, categorizing speeches into government (SDP/CDU/CSU) and opposition (Linke, Gr\"une).
}
\begin{center}
\begin{tabular}{lcccccccc}
& \multicolumn{4}{c}{\bf Held-out parliament speeches} & \multicolumn{4}{c}{\bf Party Manifestos}\\
    &         precision    &recall &  f1-score  & N    &         precision    &recall &  f1-score  & N\\
\hline \hline
government   &    0.88  &    0.95    &  0.92   &    786   &0.52   &   0.66 &     0.58  &    5972\\
 opposition    &   0.86    &  0.71   &   0.78   &    346 & 0.69  &    0.56    &  0.62   &   8229\\
\hline
avg / total    &   0.88    &  0.88     & 0.87  &    1132 &  0.62   &   0.60    &  0.60 &    14201\\
\end{tabular}
\end{center}
\end{table}

\subsection{Predicting political views}
Parties change their policies and positions in the political spectrum. More reliable categories for political bias are party independent labels for political views, see \autoref{sec:data}. A separate suite of experiments was run to train and test the prediction performance of the text classifiers models described in \autoref{sec:model}. As there was no out-of-domain evaluation set available in this setting only evaluation error on in-domain data is reported. Note however that also in this experiment the evaluation data was never seen by any model during training time.
In \autoref{tab:results_avg_political_view} results for the best and worst classes, in terms of predictability, are listed along with the average performance metrics on all classes. 
Precision and recall values of close to 0.5 on average can be considered rather high considering the large number of labels. \\

\begin{table}[t]
\caption{
\label{tab:results_avg_political_view}
Classification performance of 56 political views, see \autoref{sec:data}.
}
\begin{center}
\begin{tabular}{lcrrrr}
code & meaning  &      precision    &recall &  f1-score  & N\\
\hline\hline
       501  & environmentalism + &      0.62   &   0.61 &     0.61  &     165\\
        202 &   democracy + &   0.58  &    0.55   &   0.57   &    122\\
        701    & labour +&  0.57  &    0.54   &   0.56      & 129\\
                201    &freedom/human rights +  & 0.58   &   0.54   &   0.56   &    159\\
         106   & peace + & 0.52&      0.57   &   0.55    &    21\\
\dots\\
        302     &centralism + & 0.25     & 0.20  &    0.22  &      10\\
        401    &  free enterprise + &0.20    &  0.19    &  0.20   &     52\\
        505    &welfare - &  0.13   &   0.14   &   0.14   &     14\\
        409    & keynesian demand +&  0.14  &    0.12  &    0.13   &      8\\
            0    & undefined &  0.09  &    0.12   &   0.10    &    17\\
            \hline
avg / total  &  &  0.47    &  0.46 &     0.46 &     2946\\
\end{tabular}
\end{center}
\end{table}

\subsection{Correlations between words and parties}\label{sec:word_party_correlations}

\begin{figure}
\begin{center}
\includegraphics[width=2.8cm]{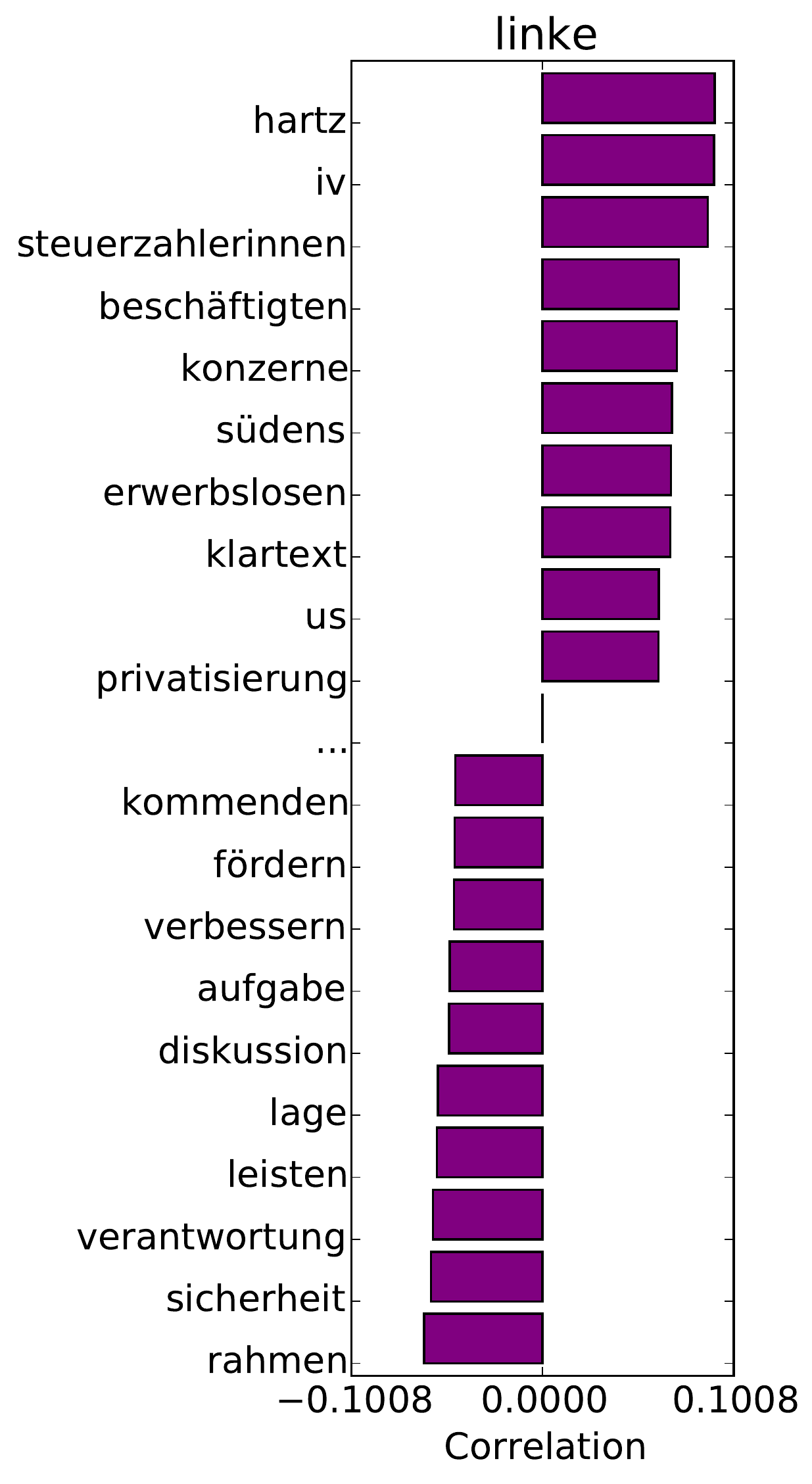} 
\includegraphics[width=2.9cm]{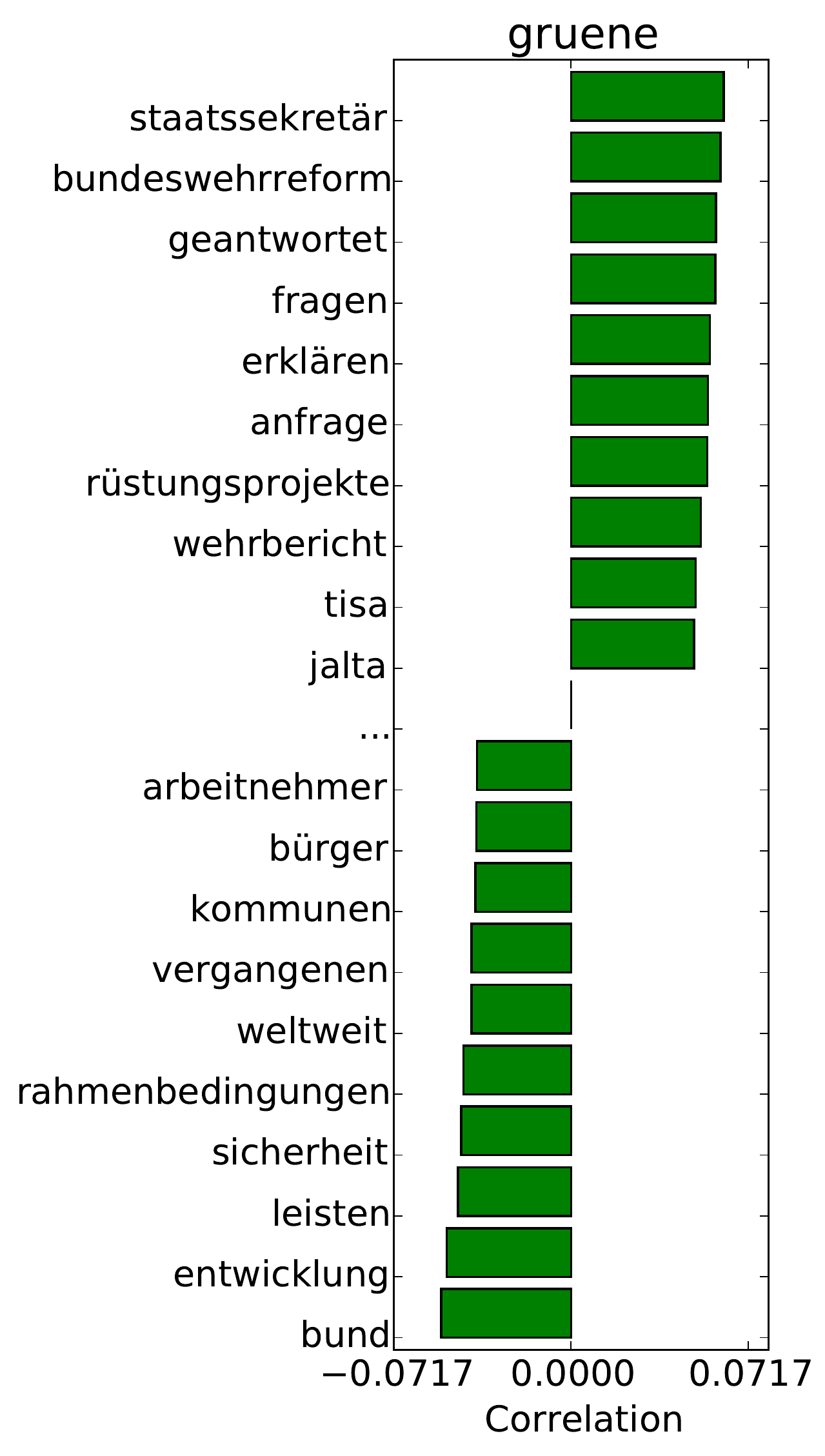} 
\includegraphics[width=3cm]{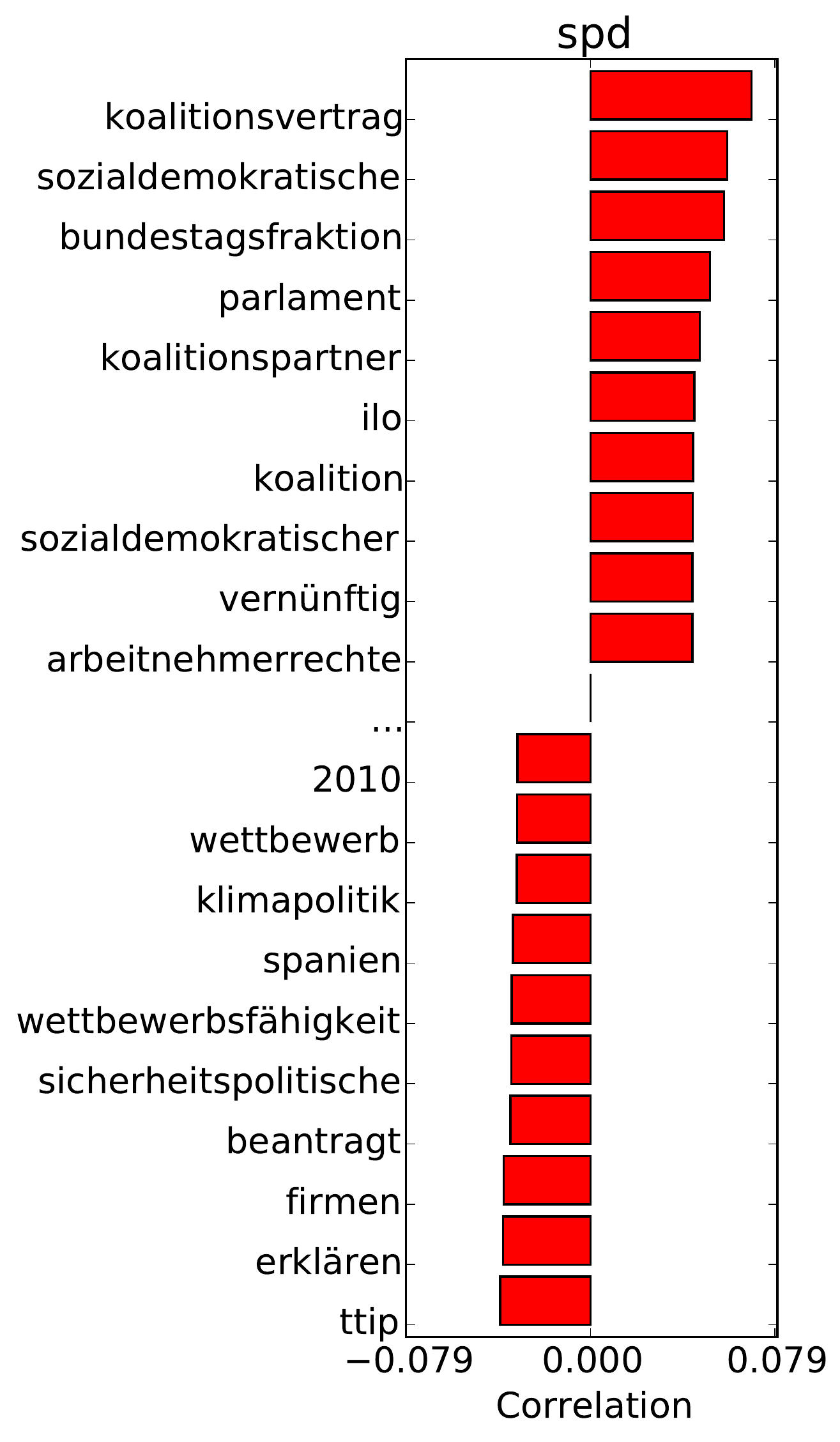} 
\includegraphics[width=3cm]{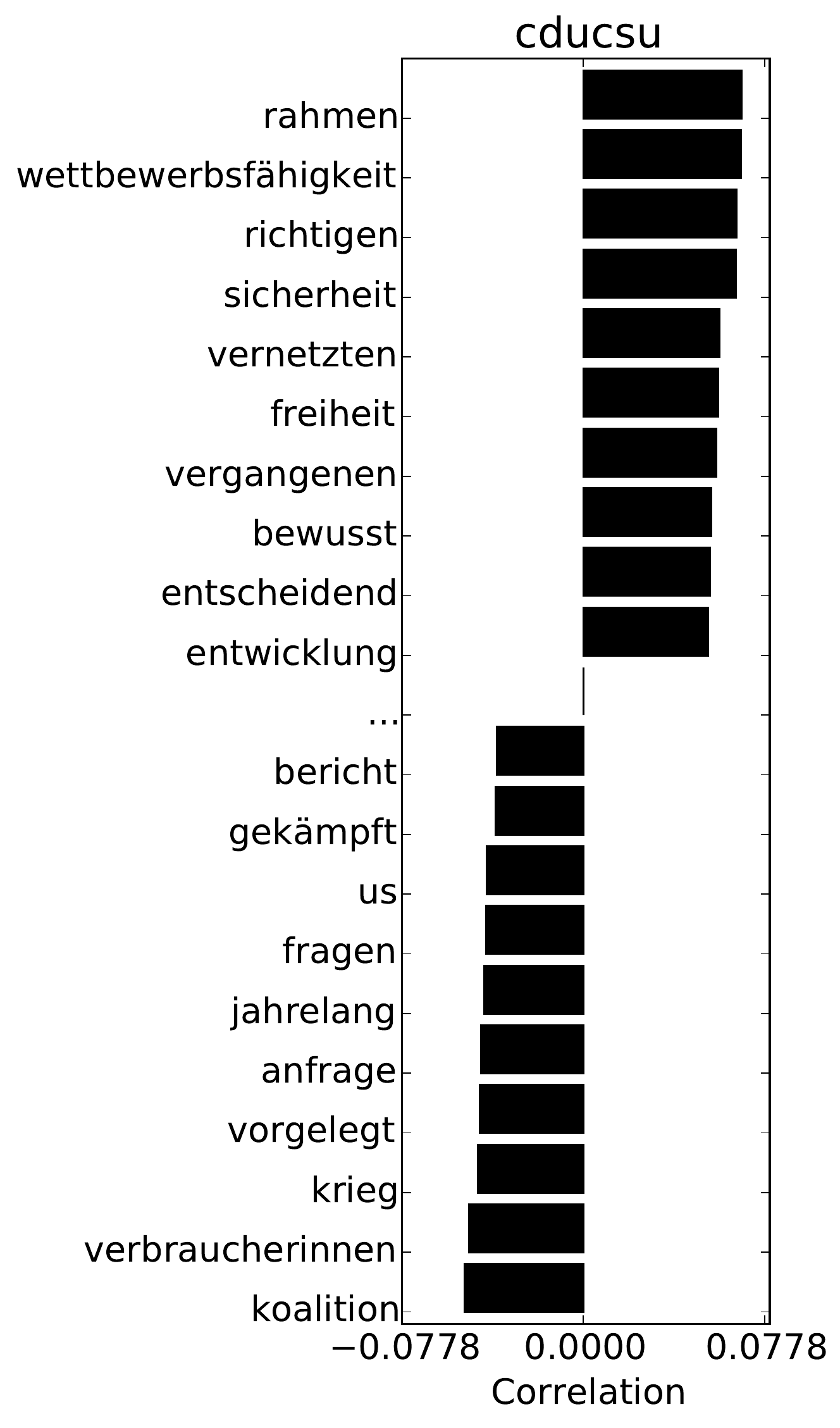}
\end{center}
\caption{
\label{fig:party_word_correlations}
Correlations between words and party affiliation label for parliament speeches can help interpreting the features used by a predictive model. Shown are the top 10 positively and negatively correlated text features for the current Bundestag. For interpretations see \autoref{sec:word_party_correlations}.}
\end{figure}

The 10 highest and lowest correlations between individual words and the party affiliation label are shown for each party in \autoref{fig:party_word_correlations}. Correlations were computed on the data from the current, 18th, legislative period. Some unspecific stopwords are excluded. 
The following paragraphs highlight some examples of words that appear to be preferentially used or avoided by each respective party. Even though interpretations of these results are problematic in that they neglect the context in which these words were mentioned some interesting patterns can be found and related to the actual policies the parties are promoting. 
\paragraph{\bf Left party (linke)}
The left party mostly criticises measures that affect social welfare negatively, such as the {\em Hartz IV} program. Main actors that are blamed for decisions of the conservative governments by the left party are big companies ({\em konzerne}). Rarely the party addresses concerns related to security ({\em sicherheit}). 
\paragraph{\bf Green party (gruene)}
The green party heavily criticised the secret negotiations about the TiSA agreement\footnote{\url{https://en.wikipedia.org/wiki/Trade_in_Services_Agreement}} and insists in formal inquiries that the representatives of the green party put forward in this matter ({\em fragen, anfragen}). They also often ask questions related to army projects ({\em R\"ustungsprojekte, Wehrbericht}) or the military development in east europe ({\em  Jalta}\footnote{Referring to the \url{https://en.wikipedia.org/wiki/Yalta_Conference}}).
\paragraph{\bf Social democratic party (SPD)}
The social democrats often use words related to rights of the working class, as reflected by the heavy use of the {\em International Labour Organisation} (ILO) or rights of employes ({\em Arbeitnehmerrechte}). They rarely talk about competition ({\em Wettbewerb}) or climate change ({\em klimapolitik}). 

\paragraph{\bf Conservative party (CDU/CSU)}
The conservative christian party often uses words related to a pro-economy attitude, such as competitiveness or (economic) development ({\em Wettbewerbsf\"ahigkeit, Entwicklung}) and words related to security ({\em Sicherheit}). The latter could be related to the ongoing debates about whether or not the governments should be allowed to collect data and thus restrict fundamental civil rights in order to better secure the population. In contrast to the parties of the opposition, the conservatives rarely mention the word war ({\em krieg}) or related words.

\subsection{Speech sentiment correlates with political power}\label{sec:sentiment_result}
\begin{figure}
\begin{center}
\includegraphics[width=6cm]{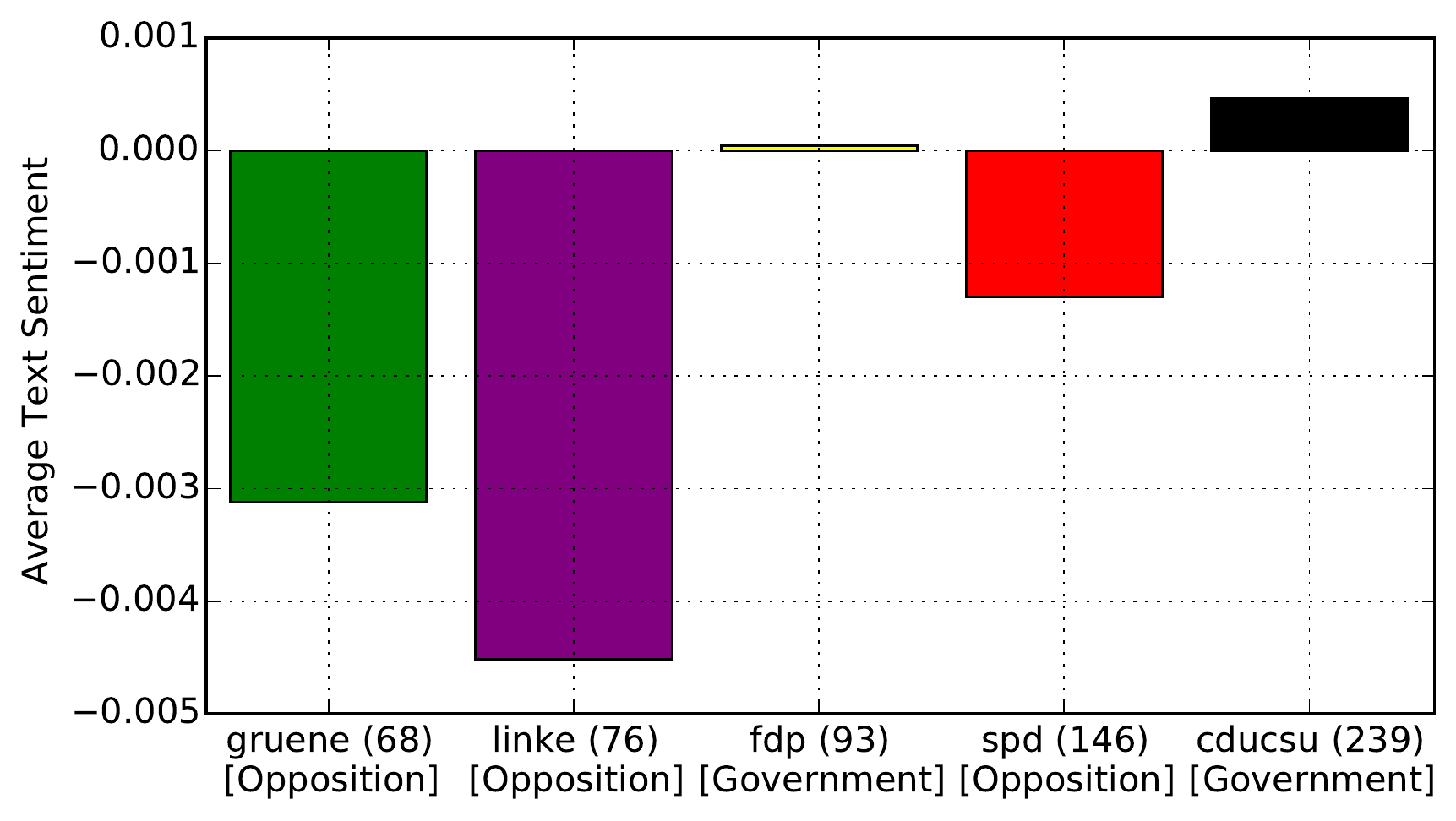}  \hfill \includegraphics[width=5cm]{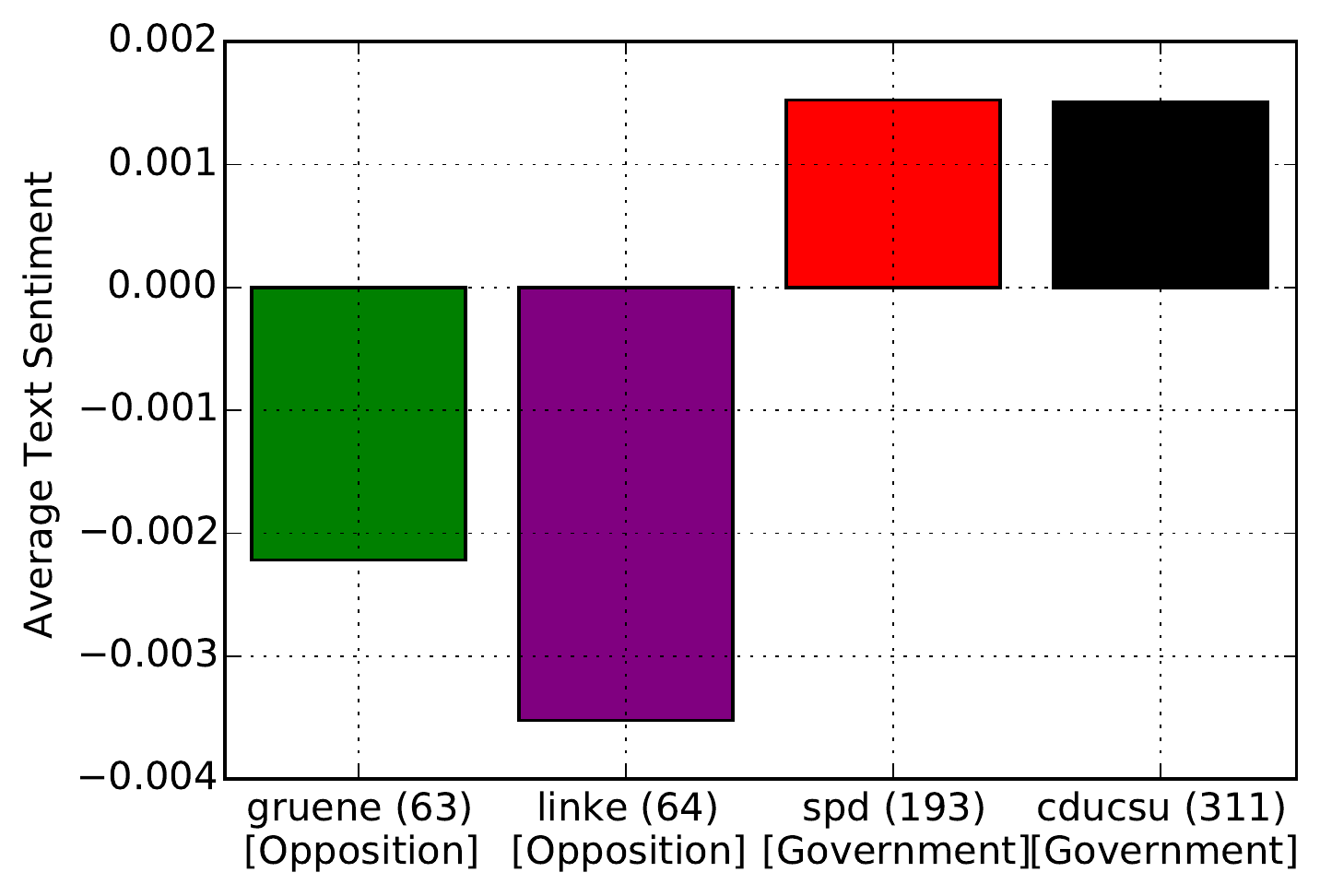} 
\end{center}
\caption{
\label{fig:party_sentiments}
Speech sentiments computed for speeches of each party; parties are ordered according to the number of seats in the parliament. There is a trend for more positive speech content with more political power. Note that the SPD (red) switched from opposition to government in the 18th Bundestag: their seats in the parliament increased and the average sentiment of their speeches switched sign from negative to overall positive sentiment.
}
\end{figure}

In order to investigate the features that give rise to the classifiers' performance the bag-of-words features were analysed with respect to their sentiment. The average sentiment of each political party is shown in \autoref{fig:party_sentiments}. High values indicate more pronounced usage of positive words, whereas negative values indicate more pronounced usage of words associated with negative emotional content. 

The results show an interesting relationship between political power and sentiment. Political power was evaluated in two ways: a) in terms of the number of seats a party has and b) in terms of membership of the government. Correlating either of these two indicators of political power with the mean sentiment of a party shows a strong positive correlation between speech sentiment and political power. This pattern is evident from the data in \autoref{fig:party_sentiments} and in \autoref{tab:sentiments}: In the current Bundestag, government membership correlates with positive sentiment with a correlation coefficient of 0.98 and the number of seats correlates with 0.89. 

Note that there is one party, the social democrats (SPD), which has many seats and switched from opposition to government with the 18th Bundestag: With its participation in the government the average sentiment of this party switched sign from negative to positive, suggesting that positive sentiment is a strong indicator of government membership.

\begin{table}[t]
\caption{
\label{tab:sentiments}
Correlation coefficient between average sentiment of political speeches of a party in the german Bundestag with two indicators of political power, a) membership in the government and b) the number of seats a party occupies in the parliament. 
}
\begin{center}
\begin{tabular}{lcc}
   Sentiment vs. &          Gov. Member    &  Seats\\
\hline\hline
17th Bundestag    &  0.84 & 0.70\\
18th Bundestag   &  0.98 & 0.89\\
\end{tabular}
\end{center}
\end{table}

\subsection{An example web application}
\begin{figure}
\begin{center}
\includegraphics[width=10cm]{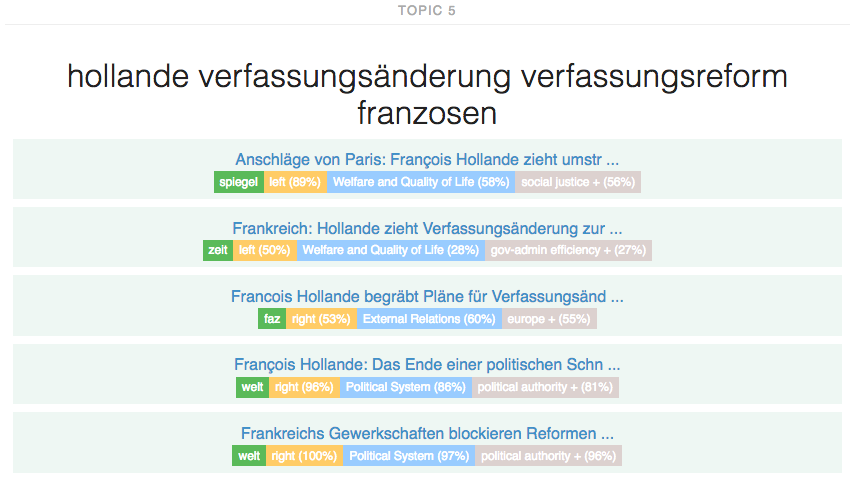}
\end{center}
\caption{
\label{fig:fipi}
A screen shot of an example web application using the political view prediction combined with topic modelling to provide a heterogeneous overview of a topic. }
\end{figure}

To show an example use case of the above models a web application was implemented that downloads regularly all articles from some major german news paper websites\footnote{\url{http://www.spiegel.de/politik}, \url{http://www.faz.net/aktuell/politik}, \url{http://www.welt.de/politik}, \url{http://www.sueddeutsche.de/politik}, \url{http://www.zeit.de/politik}} and applies some simple topic modelling to them. For each news article topic, headlines of articles are plotted along with the predictions of the political view of an article and two labels derived deterministically from the 56 class output, a left right index and the political domain of a text, see \cite{leftright}. Within each topic it is then possible to get an ordered (from left to right) overview of the articles on that topic. An example of one topic that emerged on March 31st is shown in \autoref{fig:fipi}. A preliminary demo is live at \cite{fipidemo} and the code is available on github\cite{fipi}.

\section{Conclusions, Limitations and Outlook}\label{sec:conclusion}
This study presents a simple approach for automated political bias prediction. The results of these experiments show that automated political bias prediction is possible with above chance accuracy in some cases. It is worth noting that even if the accuracies are not perfect, they are above chance and comparable with results of comparable studies \cite{Yu2008,Hirst2014}. While these results do not allow for usage in production systems for classification, it is well possible to use such a system as assistive technology for human annotators in an active learning setting.

One of the main limiting factors of an automated political bias prediction system is the availability of training data. Most training data sets that are publicly available have an inherent bias as they are sampled from a different domain. This study tried to quantify the impact of this effect. 
For the cases in which evaluation data from two domains was available there was a pronounced drop in prediction accuracy between the in domain evaluation set and the out of domain evaluation set. This effect was reported previously for similar data, see e.g. \cite{Yu2008}. Also the finding that shorter texts are more difficult to classify than longer texts is in line with previous studies \cite{Hirst2014}. When considering texts of sufficient length (for instance by aggregating all texts of a given political topic) classification performance improved and in some cases reliable predictions could be obtained even beyond the training text domain.

Some aspects of these analyses could be interesting for social science researchers; three of these are highlighted here.
First the misclassifications of a model can be related to the changes in policy of a party. Such analyses could be helpful to quantitatively investigate a change in policy. Second analysing the word-party correlations shows that some discriminative words can be related to the political views of a party; this allows for validation of the models by human experts. Third when correlating the sentiment of a speech with measures of political power there is a strong positive correlation between political power and positive sentiment. While such an insight in itself might seem not very surprising this quantifiable link between power and sentiment could be useful nonetheless: Sentiment analysis is a rather domain independent measure, it can be easily automated and scaled up to massive amounts of text data. Combining sentiment features with other measures of political bias could potentially help to alleviate some of the domain-adaptation problems encountered when applying models trained on parliament data to data from other domains. \\

All data sets used in this study were publicly available, all code for experiments and the link to a live web application can be found online \cite{fipi}.

\subsection*{Acknowledgements}
I would like to thank Friedrich Lindenberg for factoring out the \url{https://github.com/bundestag/plpr-scraper} from his bundestag project. Some backend configurations for the web application were taken from an earlier collaboration with Daniel Kirsch. Pola Lehmann and Michael Gaebler provided helpful feedback on an earlier version of the manuscript. Pola Lehman also helped with getting access to and documentation on the Manifestoproject data. 
\small{
\bibliographystyle{plain}
\bibliography{political_bias_prediction} 
}

\end{document}